\documentstyle[12pt,graphicx]{article}

\newcommand{\be}{\begin{equation}}
\newcommand{\ee}{\end{equation}}
\newcommand{\ba}{\begin{array}{l}}
\newcommand{\ea}{\end{array}}
\newcommand{\banonum}{\begin{eqnarray*}}
\newcommand{\eanonum}{\end{eqnarray*}}
\newcommand{\baa}{\begin{eqnarray}}
\newcommand{\eaa}{\end{eqnarray}}
\newcommand{\ed}{\end{document}}
\newcommand{\lab}[1]{\label{#1}}
\newcommand{\re}[1]{(\ref{#1})}
\newcommand{\ci}[1]{\cite{#1}}
\newcommand{\bfr}{\begin{flushright}}
\newcommand{\efr}{\end{flushright}}
\newcommand{\bfl}{\begin{flushleft}}
\newcommand{\efl}{\end{flushleft}}

\renewcommand{\baselinestretch}{1.4}
\date{}
\begin{document}

\title{Periodically driven dynamics of a particle moving
in the field of Coulomb plus confining potential}

\author{D.U.Matrasulov, P.K.Khabibullaev, D.M.Otajanov\\
Heat Physics Department of the Uzbek Academy of Sciences,
\\ 28 Katartal St.,700135 Tashkent, Uzbekistan\\
and\\
F.C.Khanna\\
Physics Department University of Alberta\\
Edmonton Alberta, T6G 2J1 Canada\\}

\maketitle

\begin{abstract}
Periodically driven dynamics of a particle moving in the field
Coulomb plus confining potential is treated for one and three
dimensional cases. Critical value of the external field strength
at which chaotization will occur is evaluated analytically based
on the resonance overlap criterion. The analysis of the
phase-space dynamics is presented.
\end{abstract}

\section{Introduction}

Periodically driven dynamics has been subject of extensive research
for past few decades both in classical and quantum contexts
\ci{chir}-\ci{Izr90}. An interesting feature of the periodically
perturbed dynamical system is the chaotization of its motion
under certain conditions. This implies
divergence of the phase space trajectories and unlimited energy
growth(in the classical case) leading to acceleration. Qualitative
and quantitative analysis of the model systems such as kicked
rotor shows that there is linear time dependence in the average
energy of the periodically driven system leads to a diffusion in
the phase space \ci{Izr90}.
Corresponding quantum studies have
 shown that such a diffusion is suppressed
in the quantum case \ci{Izr90}. Therefore periodically driven
systems are convenient  testing ground for the study of dynamical
chaos in time-dependent case. Great variety of periodically driven
systems have been treated from the viewpoint of dynamical chaos
theory from model systems such as kicked rotor \ci{Izr90}, pendula
\ci{esca85} or billiards \ci{eckh88}  to  realistic systems like
atoms and molecules \ci{zas88,jens,koch,del83,cas87,mat1} and
surface-state electrons in liquid helium \ci{jens84} in the
time-periodic fields. Comprehensive theoretical and experimental
studies of the hydrogen atom in a monochromatic field showed that
diffusive excitation of the atom leading to  ionization during
relatively long time can occur. This fact makes hydrogen atom in a
monochromatic field a convenient system for the theoretical and
experimental study of chaos in the time-dependent dynamical
systems \ci{cas87,jens,koch,del83}. A theoretical analysis, of the
behavior of a classical hydrogen atom interacting with
monochromatic field, based on the resonance overlap criterion
\ci{del83}-\ci{mat1}, shows that for some critical value of the
external field strength $\epsilon_{cr}$, the electron enters into
chaotic regime of motion, marked by unlimited diffusion along
orbits, leading to ionization. Experimentally, this phenomenon was
first observed  by Bayfield and Koch \ci{bay74}. Such an
ionization was called chaotic \ci{cas87,jens} or diffusive
\ci{del83} ionization. During the last three decades  chaotic
ionization of nonrelativistic atom was investigated by many
authors theoretically \ci{del83,cas87,jens84} as well as
experimentally \ci{jens,koch,bay74}.

In this paper we study a model which may be considered as a QCD
counterpart of the periodically driven hydrogen atom. Namely, we
address the problem of regular and chaotic motion of a particle
bounded in the field of Coulomb plus confining potential in the presence
of a time-periodic external perturbation. Such
system may be used for modelling periodically driven
quark-antiquark  state, so-called quarkonium.

Using resonance analysis based
on the Chirikov criterion for stochasticity we
estimate critical values of the external field strength at which quarkonium
motion enters into chaotic regime.

Quarkonium in a monochromatic field can be considered as an analog of the
hydrogen atom in a monochromatic field, in which
Coulomb potential is replaced by Coulomb plus confining potential.
Quarkonia   have been the subject of extensive experimental as well as
theoretical studies for the last two decades \ci{AL}-\ci{Mukh93}.
In the framework of potential model the description of quark motion in
hadrons is reduced to solving
classical or quantum mechanical equations of motion  with Coulomb plus
confining potential \ci{Mukh93}.

Study of the nonlinear dynamics of hadrons is of importance due to
the recent advances made in  creation of hadronic and quark-gluon
matters in the collisions experiments of ultrarelativistic heavy
ions where creation of hadronic or quark-gluon matter is possible
\ci{NA50}. Quarkonia in quark-gluon matter can be considered  as a
system perturbed by time-dependent force, that may lead to
chaotization of the quarkonium motion. Indeed, the recent studies
on quarkonium dynamics showed that regular motion can be expected
at a small values of color screening mass but the chaotic motion
is expected at a large one \ci{Gu}. Periodically driven quarkonium
can be also realized in the interaction of mesons with laser
fields. This paper is organized as follows. In next section we
will treat a simple model, a one-dimensional quarkonium under a
periodic perturbation. In section 3 we extend our treatment to the
three-dimensional case. Some concluding remarks are presented in
section 4.

\section{One-dimensional model}

For  simplicity we consider first  a one-dimensional model described by a potential
$$
V(x)=\left\{\begin{array}{ll}-\frac{Z}{x}+\lambda x\, & for\;\:x>0 \\ \\
\infty \, & for\:\;x \leq 0\end{array} \right.\,, $$

where $Z = \frac{4}{3}\alpha_{s}$, $\alpha_s$ being the effective strong coupling constant
and $\lambda$ gives strength of the confining potential.
As is well known, in the case of the hydrogen atom interacting with a monochromatic field,
one-dimensional model provides an excellent description of the experimental chaotization
thresholds for real three-dimensional hydrogen atom \ci{jens,jens84,cas87}. The same success is to be expected
in the case of quarkonium.

The unperturbed Hamiltonian for the above potential is
\be
H_0 = \frac{p^2}{2} -\frac{Z}{x} +\lambda x .
\lab{unper}
\ee

We will treat the interaction of the system given by Hamiltonian \re{unper} with the periodic external potential of the form
\be
U(x,t) = \epsilon x cos\omega t.
\lab{pert}
\ee
with $\epsilon$ and $\omega$ being the field strength and frequency, respectively.
Thus the total Hamiltonian of the periodically driven quarkonium is
\be
H =H_0 + U(x,t)
\lab{total}
\ee

Formally, the Hamiltonian \re{unper}  is equivalent to that of the
hydrogen atom in constant homogenous electric field. Chaotic
dynamics of hydrogen atom in constant electric field under the
influence of time-periodic field was treated earlier
\ci{ber,bala}. To explore periodically driven dynamics of our
system we  introduce so-called   action-angle variables and
rewrite its Hamiltonian in terms of these variables.

The action variable is defined as \ci{zas88}: \be n
=\frac{1}{2\pi} \int\limits_{c}^{a}\sqrt{2(H_0-V(x))} dx =
\frac{\sqrt{2\lambda}}{2\pi} \int\limits_{c}^{a}
\sqrt{\frac{(a-x)(x-c)}{x}}dx, \ee where the constants $a$ and $c$
are the turning points of particle and are given  by
$$
a = \frac{H_0+\sqrt{H_0^2+4Z\lambda}}{2\lambda},\;\;\;
c = \frac{H_0-\sqrt{H_0^2+4Z\lambda}}{2\lambda}
$$
Since $c<0$ for the action we have
$$
n = \frac{1}{2\pi}\int\limits_{0}^{a}\sqrt{2(H_0-V(x))} dx =
$$
\be =
B\sqrt{a+\frac{1}{a}}\cdot\Biggl[\biggl(a-\frac{1}{a}\biggr)E(k)`+
\frac{1}{a} K(k)\Biggr]. \lab{act1} \ee where
$$
B= \frac{2\sqrt{2}}{3\pi\lambda^\frac{1}{4}},
$$
here $E(k)$ and $K(k)$ are the elliptic integrals \ci{abr} and

\be
k^2 = \frac{a^2}{a^2+1}.
\lab{par}
\ee

The angle variable is defined as \ci{zas88}
$$
\theta  = \frac{\partial S}{\partial n}
$$
where
$$
S = \int^x p(x,H(n)) dx
$$
Then in terms of $n$ and $x$ variables the angle can be written as
$$
\theta(x,n) = $$
\be
B\sqrt{x+\frac{1}{x}}[(x+\frac{1}{x})\frac{E(k)-K(k)}{k}\frac{dk}{dn}
 +\frac{1}{x}K(k)
\frac{E(k)-k\sqrt{1-k^2}K(k)}{k\sqrt{1-k^2}}\frac{dk}{dn}],
\lab{eq1}
\ee
where $k =k(n)$ is defined by the eq. \re{par}.

Furthermore, we consider the following two cases: $a \gg 1$ and $a
\ll 1$. For both cases we obtain approximate expression for
unperturbed Hamiltonian as a function of angle variable. To do
this we we solve the eq.\re{act1} with respect to $H_0$ and use
asymptotic estimates for the elliptic integrals, $E(k)$ and $K(k)$
\ci{abr} for $a<< 1$ and $a>> 1$ that correspond to $k <<1$ and $k
\approx 1$, respectively.

For $a\gg 1$ we have:
\be
H_0 =
 Z^2An^{2/3} \cdot\Biggl[1 - \frac{\lambda ln(4B^{-2/3}n^{2/3})}{A^2n^{4/3}} \Biggr],
\ee
with
$$
A= \frac{3\pi \lambda}{2\sqrt{2}}^\frac{2}{3}.
$$

Corresponding proper frequency is
\be
\omega_0 = \frac{2}{3}Z^2 \biggr[
\frac{A}{n^{1/3}} + \frac{\lambda}{An^{5/3}}\bigr[\ln(4A\sqrt{\lambda}n^{2/3})\bigl]-1 \biggl]
\lab{freq}
\ee

For second case, $a\ll 1$ we get
\be
H_0 = 0.5 Z^2 (9.7\lambda n^2 - n^{-2});
\ee

The proper frequency for this Hamiltonian is
\be
\omega_0 = Z^2 (n^{-3}+9.7n\lambda).
\ee

The Hamiltonian of the perturbed system can be written as
\be
H = H_0 + \epsilon \sum x_k cos(k\theta -\omega t),
\lab{full}
\ee

with

\be
x_k(n) = - \int_{0}^{2\pi}x(n,\theta) e^{ik\theta}d\theta
\ee
being Fourier amplitude  of the  perturbation.
For $a\ll 1$ we have
\be
x_k(n)\approx-\frac{4E(n)}{\lambda}
\frac{1}{k}
\sin^2\frac{\pi k \sqrt{\lambda}}{2}.
\ee
For  $a\gg 1$ we obtain
\be
x_k(n)=-\frac{2An^{2/3}}{\pi^2\lambda k^2}.
\ee

As is well known \ci{chir,esca85,zas88}, dynamics of a periodically driven system may become chaotic if the
resonance condition is fulfilled and external field strength exceeds some critical value.
To estimate this critical field strength, $\epsilon_{cr}$ for our system
we use Chirkov's resonance overlap criterion
\ci{chir,zas88,del83}, which can be written as:
\be
\frac{\Delta \nu_k +\Delta
\nu_{k+1}}{\omega_{0}(k+1)-\omega_{0}(k)} >2.5,
\lab{chir} \ee with
$$
\Delta \nu_k = (\frac{\epsilon x_k}{\omega_0'})
$$
being the width of the $k$-th resonance \ci{chir,zas88}
and
$$
\omega_0' = d\omega_0/dn.
$$
From the resonance condition we get
$$
{\omega_{0}(k)-\omega_{0}(k+1)} = \frac{\omega}{k}-\frac{\omega}{k+1} =
\frac{\omega}{k(k+1)}.
$$
Applying this criterion to our system give by  \re{full} we have for $a \gg1$

$$
\epsilon_{cr} = \frac{ 0.07 Z^2 \omega \pi^2 \lambda}{ n^2}
\cdot \frac{k (k+1)}{(k+1)^2 + k^2} \times
$$
\be
\left\{
 1  + \frac{ \lambda }{ A^2 n^\frac{4}{3} }
\left[ 5 \ln \left(  4 A \lambda^{-\frac{1}{2}}
n^\frac{2}{3} \right) - 7 \right]
\right\}
\lab{larg}
\ee

and for  $a\ll 1$:

$$
\epsilon_{cr} = \frac{0.3 \omega \lambda}{k(k+1) n^2}\cdot \frac{29\lambda
n^4-9}{29\lambda n^4-3}\times
$$
\be
\times \left[ \frac{1}{k} \sin^2 ( k \sqrt{\lambda} \frac{\pi}{2})+
\frac{1}{k+1}  \sin^2 ( (k+1) \sqrt{\lambda} \frac{\pi}{2})
\right]^{-1}
\lab{smal}
\ee

Table 1 presents the values of the critical field for $a<<1$
$u\bar u$, $d\bar d$, $s\bar s$, $c\bar c$ and $b\bar b$ quarkonia
at the following values of parameters: $\omega =10^{9}Hz$,
$\alpha_s =0.112$\ci{PDG}, $\lambda =0.2GeV^2$, $n=5;7;10$. For
light ($u,d,s$) quarkonia we use formula \re{larg} and formula
\re{smal} for $b\bar b$- and $c\bar c$- quarkonia. In the Table 2
the critical values of the field strength are displayed for the
same values of parameters but at $a>>1$. The huge difference
between the corresponding values in these two tables can be
explained by the fact that these two limit cases make our system
(to some extent) equivalent to hydrogenlike atom($a<<1$) and
highly excited confined(by linear potential only) system.
Therefore the difference between these two cases is caused by the
big difference between the binding energies.

In Fig.1 the critical field strength is plotted as a function of the
action n. Again, three cases are considered: periodically driven
motion in the field Coulomb potential ($\lambda =0$), periodically
driven motion in the field of Coulomb plus linear potential for
the cases $a \ll 1$ and for $a\gg 1$.

The parameters $Z$ and $\lambda$ are chosen as: $Z=0.15$; $\lambda
= 0.4$. As is seen from this figure, unlike to periodically driven
hydrogen-like atom, where field strength is a decreasing function
of n, for periodically driven quarkonium field strength increases
first up to certain value of n, then decreasing occurs. The curve
for quarkonium at $a<<1$ is closer to that for hydrogen-like atom,
while for $a>>1$ the difference between the critical fields for
quarkonium and hydrogen-like atom becomes significant. This can be
explained by the fact that for $a<<1$ the  energy of the
unperturbed quarkonium becomes closer to that for hydrogen-like
atom.

It  is well known that  the phase-space trajectories of the
regular motion lie on tori(so-called KAM tori). According to
Kolmogorov-Arnold-Moser theorem for sufficiently  small fields
most of the trajectories remain regular. If the value of the
external perturbation exceeds  some value, which is called the
critical field strength, KAM tori  start to break down and
chaotization of the motion will occur \ci{zas88}.

In Figs 2-4 the phase-space portraits are plotted and compared for
periodically driven pure Coulomb potential (hydrogenlike atom),
quarkonium system at $a \ll 1$ and quarkonium system at $a\gg 1$
for various values of the ratio $\epsilon_{cr}/\epsilon$.

Fig. 2 presents phase-space portraits the case $ \epsilon_{cr}/
\epsilon = 0.1$; the panels $a, b, c$ are the phase space
portraits at hydrogenlike atom, quarkonium for $a \ll 1$ and
quarkonium for $a \gg 1$, respectively. Figs. 3 and 4 correspond
to the cases when $\epsilon_{cr}/\epsilon = 0.5$ and
$\epsilon_{cr}/\epsilon = 0.9$, respectively.

The following values of the parameters $Z, \lambda$ were chosen
for the plots: $Z=0.15$; $\lambda = 0.4$.

As is seen from these plots periodically driven motion in the
field of Coulomb potential is more chaotic than that in the field
of Coulomb plus linear potential. In other words confining
potential makes periodically driven dynamics more regular. Among
two limit cases, $a \ll 1$ and $a \gg 1$, dynamics for $a \ll 1$
is more chaotic than that for $a \gg 1$, which also confirms that
confining force suppresses chaotization of motion.

\section{Three-dimensional model}

The Hamiltonian for the three-dimensional model is
$$
H_0 = \frac{p^2_r}{2} - \frac{Z}{r} +\lambda r +\frac{L^2}{r^2}.
$$
where $L$ is the orbital angular momentum and $p_r$ is the radial momentum.

The action can be expressed in terms of  elliptic integrals \ci{seet}:
$$
n = \int\limits_{c}^{a} p dr = \int\limits_{c}^{a}
\sqrt{2(E
- \frac{L^2}{r^2}+ \frac{Z}{r} -\lambda r )} dr=
$$
$$\left[
(2Z/3 -L^2/c +Ec/3) K(k) +E(a-c)/3 E(k) + \right.
$$
\be
\left.+ L^2 (c^{-1}-b^{-1}) \prod(\beta^2 , k)
\right]g/\sqrt{\lambda}
\lab{act2}
\ee

with $a$ and $c$ being  the turning points,
$K,E,\prod$ are complete elliptic integrals of the first, second and third kind,
respectively \ci{abr}, and the constants are given as
$$
k^2 = (a-b)/(a-c),
$$
$$
\beta = c k^2 / b,
$$
$$
g=2/\sqrt{a-c}.
$$
From eq.\re{act2}  the unperturbed Hamiltonian as a function of $n$ can be found approximately for $E/\lambda \gg1$,
(which corresponds to $n \gg 1$
or large quarkonium masses):
$$
H_0=(\frac{3}{2} \lambda n)^{2/3}\left[1+\frac{\pi L}{3 n} \right].
$$
The proper frequency is
$$
\omega_0 = \frac{\partial H_0}{\partial n} =(\frac{2\lambda^2}{3})^{1/3}.
\left[
n^{-1/3} -\frac{\pi L}{6}n^{-4/3}
\right]
$$

Then the Hamiltonian of the three-dimensional quarkonium in a
monochromatic field can be written as
$$
H=H_0 + \epsilon a \cos (\omega t) \times
$$
\be
\{ -\frac{3}{2} e \sin \varphi +
2 \sum [ x_k \sin\psi \cos k\phi + y_k \cos\psi \sin k\phi ] \},
\ee

where
$$
x_k = \frac{2i}{\omega_0 kT}\int e^{i\omega_0 kt}\dot x dt, \;\;\;\;y_k = \frac{2i}{\omega_0 kT}\int e^{i\omega_0 kt}\dot y dt,
$$
and $\psi$ and $\phi$ are the Euler angles.
Again, using the  resonance overlap criterion \re{chir} in which the resonance width is defined by
$$
\Delta \nu_k = (\frac{\epsilon r_k}{\omega_0'}),
$$
where
$$
r_k =\sqrt{x_k^2+y_k^2},
$$
we obtain an estimate for the critical field:

$$
\epsilon_{cr}= \frac{0.07 \lambda\omega }{ k(k+1)\pi n^2 }
\left( 1 - \frac{\pi L }{n} \right)
$$
$$
\left\{
\sqrt{  \frac{16 \pi^2}{9}+\frac{1}{k^2} }
+ \sqrt{ \frac{16 \pi^2}{9}+\frac{1}{(k+1)^2} }
\right\}^{-1}
$$
\be
\times
\left[ 1 - \frac{L^2 }{4 \pi^4 n^2}  \right].
\ee

This estimate for the critical field assumes that $n >>1$.
If the external field strength has  the value exceeding $\epsilon_{cr}$, breaking of KAM surfaces in the pase space will occur and
and the quarkonium diffuses in action and  the motion becomes chaotic.

\section{Conclusions}

In this work we explored periodically driven dynamics of a
particle bounded in the field of Coulomb plus linear potential.
Using resonance overlap criterion the estimates of for the
critical field strength at which dynamics enters into chaotic
regime of motion are obtained. The analysis of phase-space
dynamics by plotting phase space portraits shows that the
periodically driven dynamics of quarkonium is more regular
compared to that for hydrogen-like atom. Also, there appears a
pick in n-dependence of the critical field strength, that can be
explained by the role of confining potential. The obtained results
are applied for the estimation of critical field strength needed
for chaotization of various quarkonia. The model considered in
this work can be used for modelling nonlinear dynamics of
quarkonia perturbed by the field of a quark-gluon plasma or laser
radiation.

\vskip0.5cm

\section{Acknowledgements}

This work is supported by the NATO reintegration grant (Ref. No.
981787), by INTAS Young Scientist fellowship (Ref. No.
06-1000023-6008) and by a grant of the Uzbek Academy of Sciences
(FA-F2-084). The work of FCK is supported by NSERCC.

\newpage

\begin{center}

TABLE 1. The values of the critical field strength for heavy
quarkonia.
 \vskip 0.5 cm

\begin{tabular}{|c|c|c|c|c|c|c|} \hline
\multicolumn{1}{|c|}{No}& \multicolumn{1}{|c|}{Quarkonium} &
\multicolumn{1}{|c|}{ Quark mass (in MeV)} &
\multicolumn{3}{|c|}{ Critical field (V/fm)}\\
\cline{4-6}
&&&$n=5$ & $n=7$  & $n=10$ \\
\hline
1 & $c\bar c$ & $300$ & $1.215$& $0.6163$ & $0.3008$ \\
\hline
2 & $b\bar b$ & $1560$ & $5.761 \cdot 10^2 $& $2.901\cdot 10^2$ & $1.407\cdot 10^2$ \\
\hline
\end{tabular}

\vskip 2 cm

TABLE 2. The values of the critical field strength for light
quarkonia.
 \vskip 0.5 cm

\begin{tabular}{|c|c|c|c|c|c|c|} \hline
\multicolumn{1}{|c|}{No}& \multicolumn{1}{|c|}{Quarkonium} &
\multicolumn{1}{|c|}{ Quark mass (in MeV)} &
\multicolumn{3}{|c|}{ Critical field (V/fm)}\\
\cline{4-6}
&&&$n=5$ & $n=7$  & $n=10$ \\
\hline
1 & $u\bar u$ & $1 $ & $1.018\cdot 10^{19}$& $5.192\cdot 10^{18}$ & $2.544\cdot 10^{18}$\\
\hline
2 & $d\bar d$ & $2$ & $ 8.141\cdot 10^{19}$& $4.153\cdot 10^{19}$ & $2.035\cdot 10^{19}$ \\
\hline
3 & $s\bar s$ & $30$ & $9.158\cdot 10^{22}$& $4.673\cdot 10^{22}$ & $2.29\cdot 10^{22}$ \\
\hline
\end{tabular}

\end{center}

\newpage

\begin{figure}[t]
\begin{center}
\includegraphics{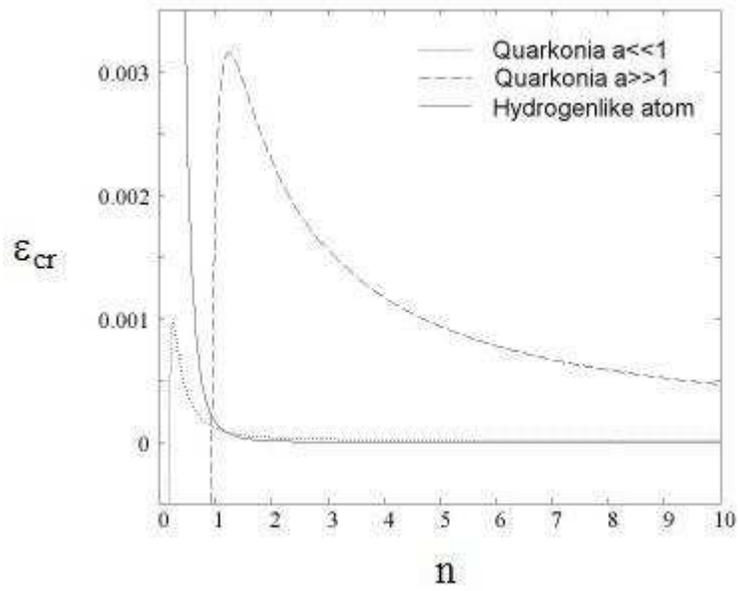}

\vspace{.5cm} \caption{  Critical field strength (in the system of
units where quarkonium mass is equal to 1) as a function of n, for
periodically driven hydrogenlike atom, quarkonium in $a \gg 1$
case and quarkonium in $a \ll 1$ cases. }
\end{center}
\end{figure}

\newpage

\begin{figure}[t]
\begin{center}
\includegraphics[width=8cm]{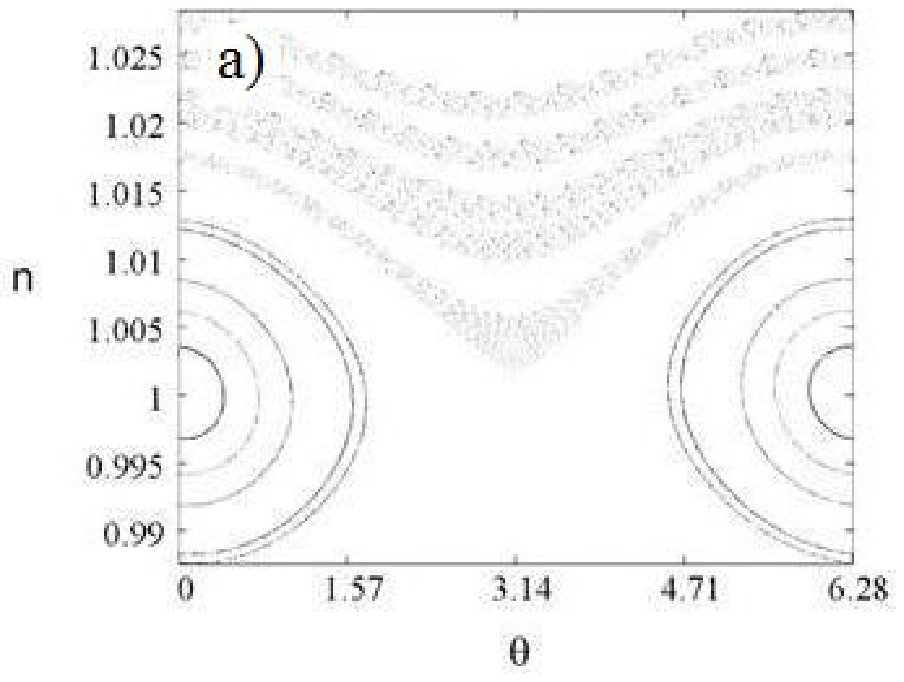}
\includegraphics[width=8cm]{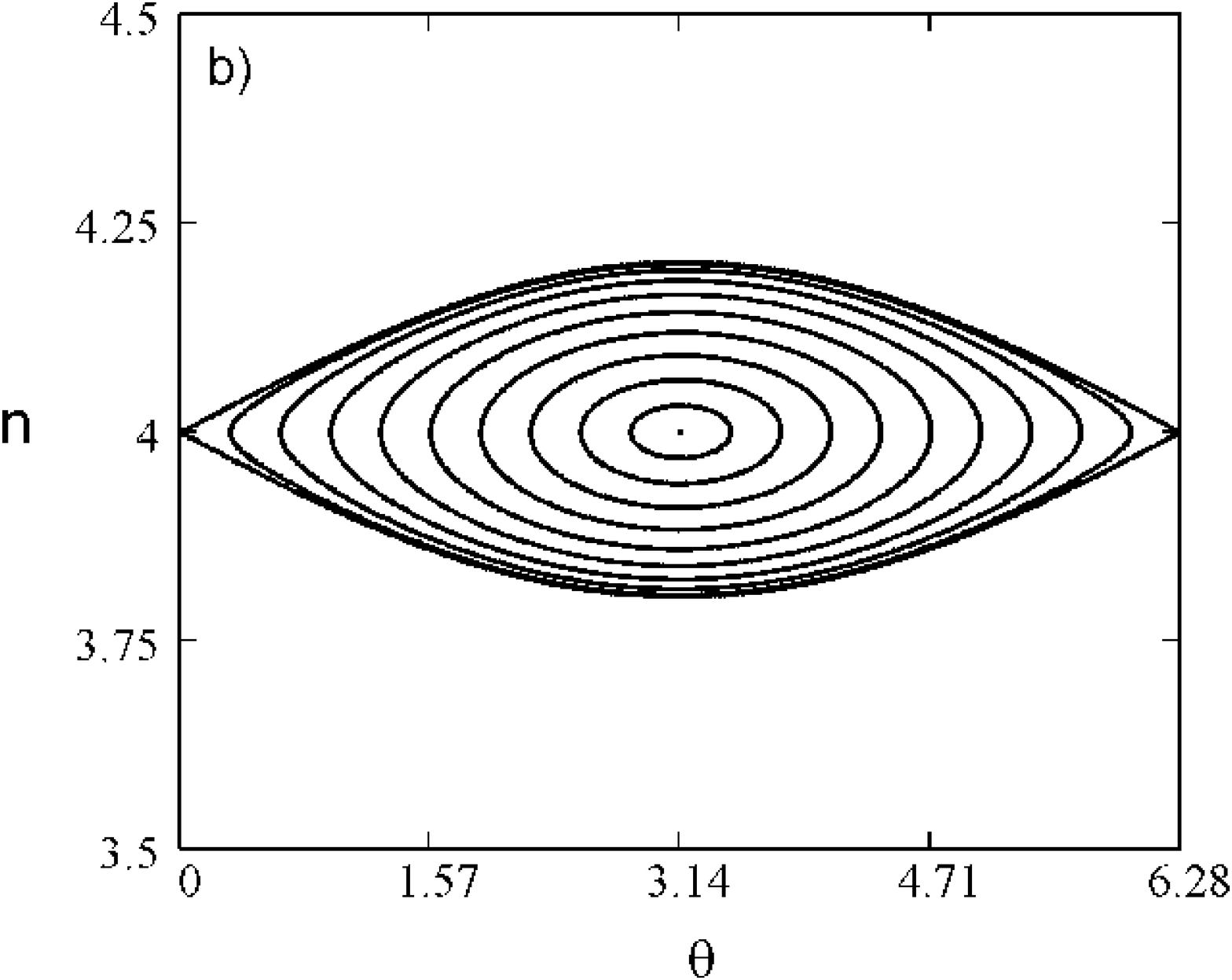}
\includegraphics[width=8cm]{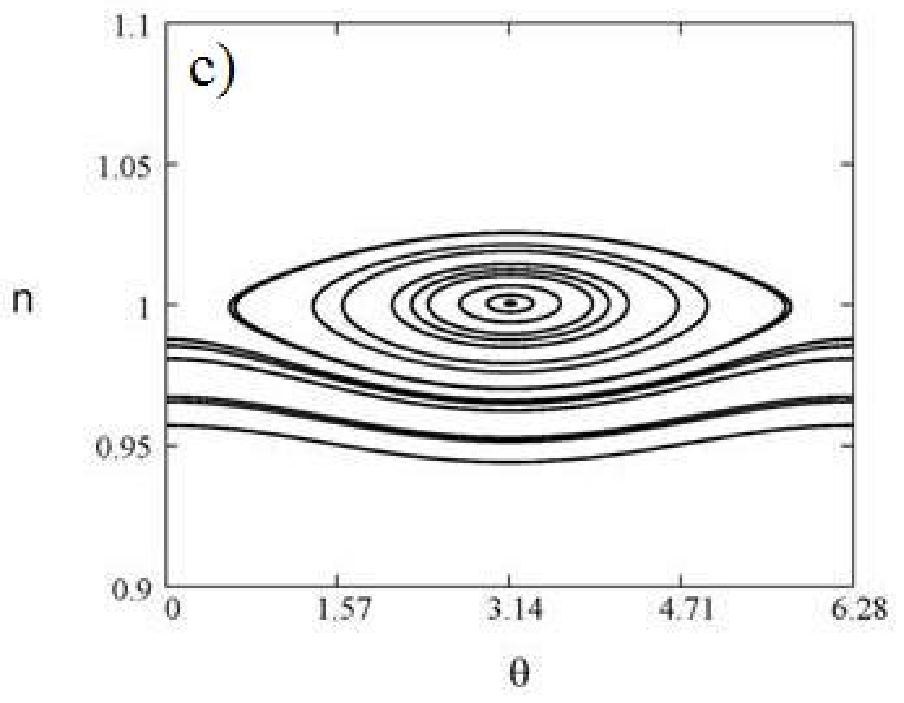}
\vspace{.5cm} \caption{  Phase space portraits of the periodically
driven: a) hydrogenlike atom; b) quarkonium ($a \gg 1$ case); c)
quarkonium ($a \ll 1$ case), for $\epsilon_{cr}/\epsilon = 0.1$.}
\end{center}
\end{figure}

\newpage

\begin{figure}[t]
\begin{center}
\includegraphics[width=8cm]{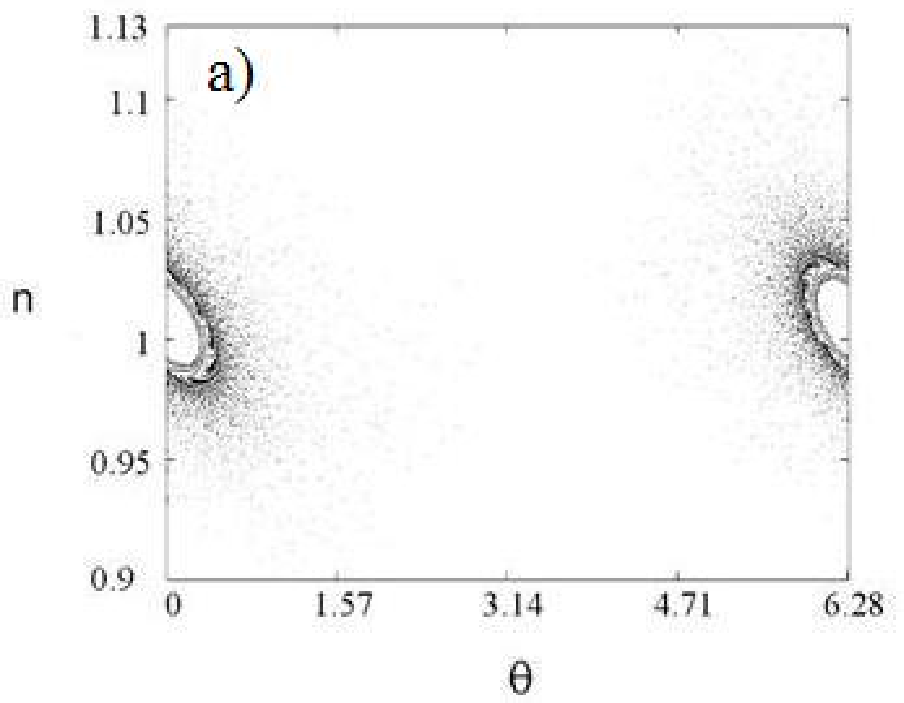}
\includegraphics[width=8cm]{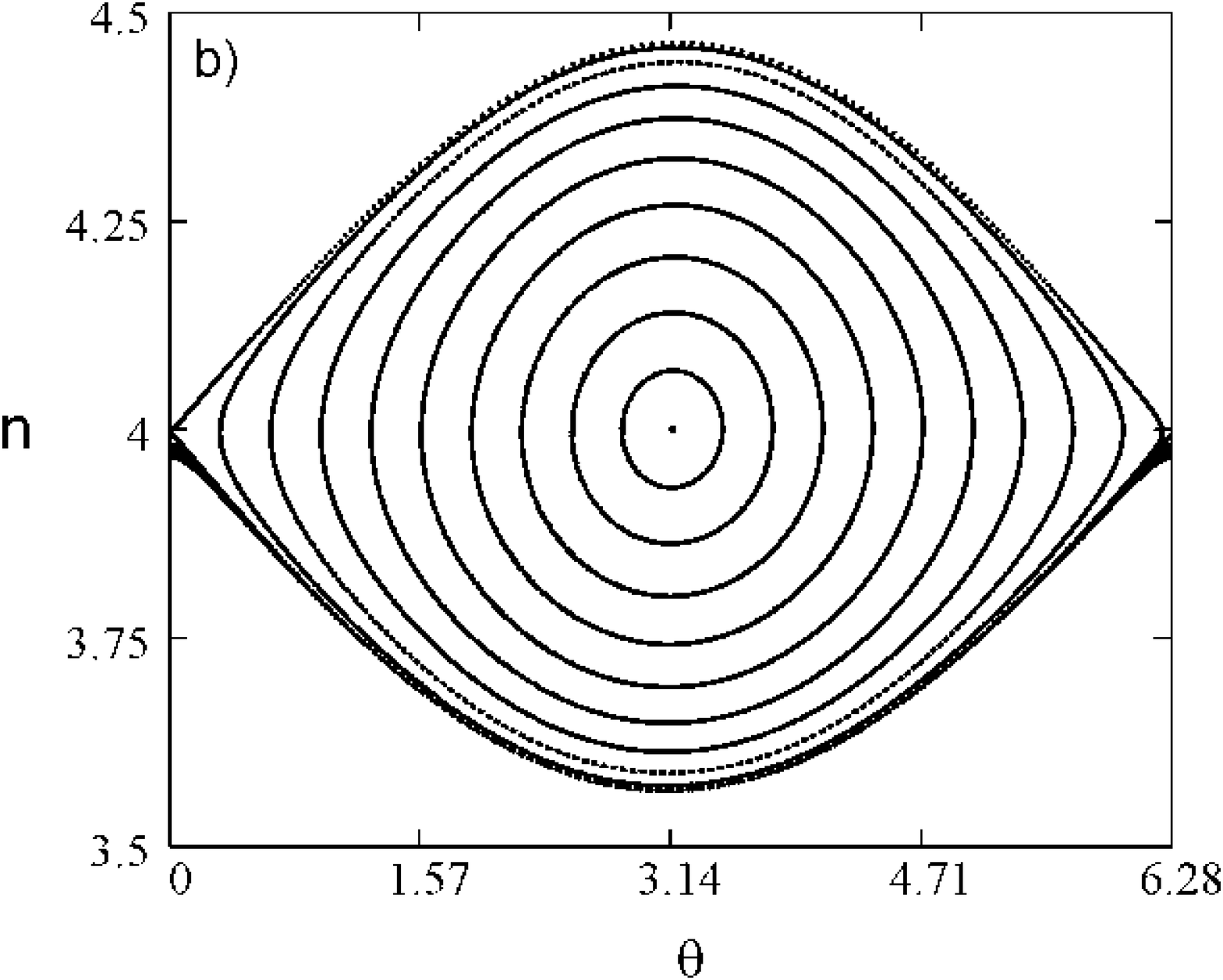}
\includegraphics[width=8cm]{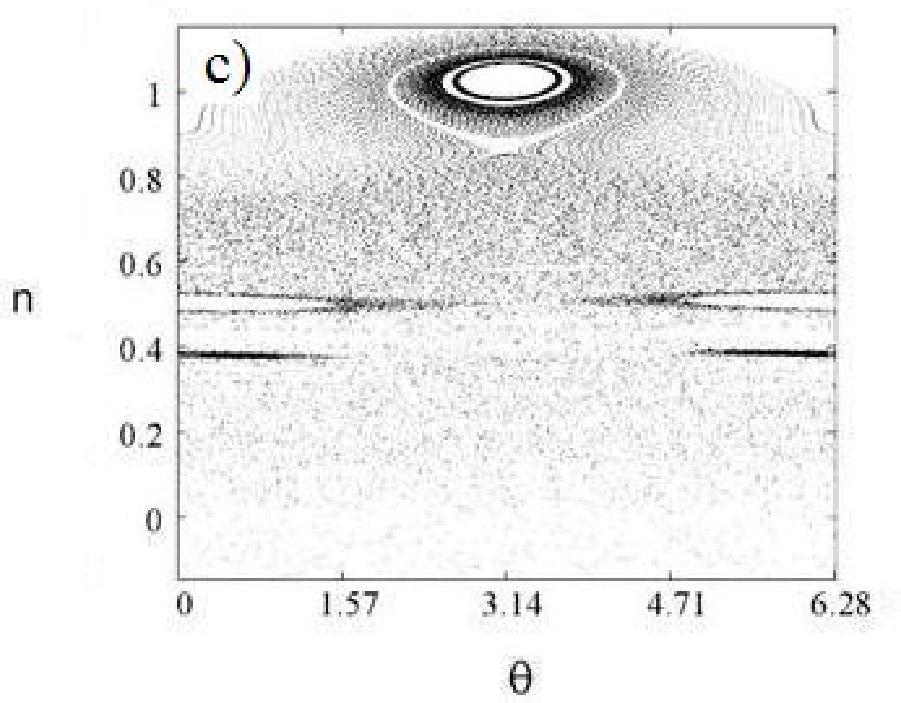}

\vspace{.5cm} \caption{  Phase space portraits of the periodically
driven: a) hydrogenlike atom ; b) quarkonium ($a \gg 1$ case); c)
quarkonium ($a \ll 1$ case), for $\epsilon_{cr}/\epsilon = 0.5$.}
\end{center}
\end{figure}

\newpage

\begin{figure}[t]
\begin{center}
\includegraphics[width=8cm]{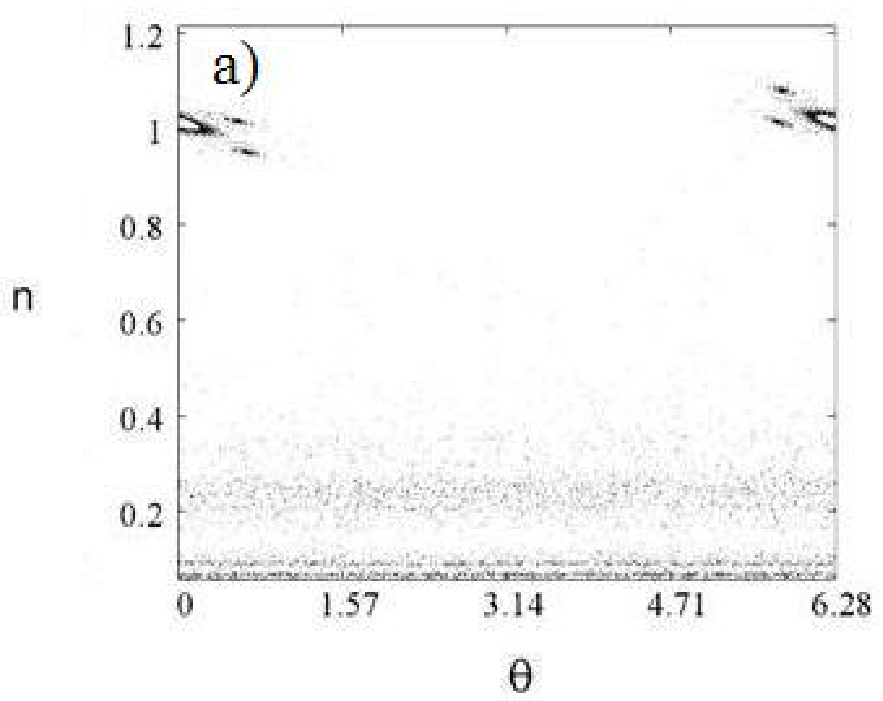}
\includegraphics[width=8cm]{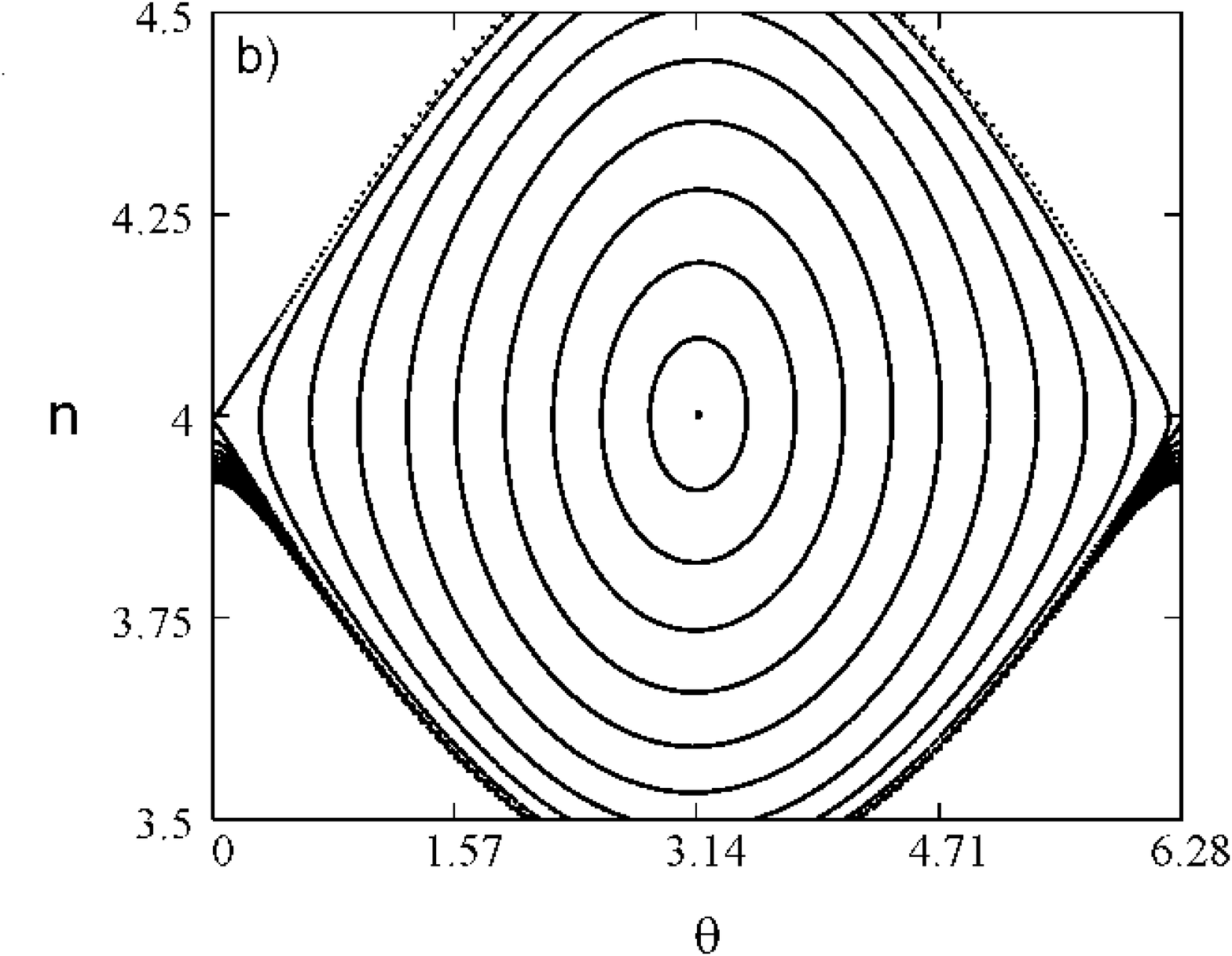}
\includegraphics[width=8cm]{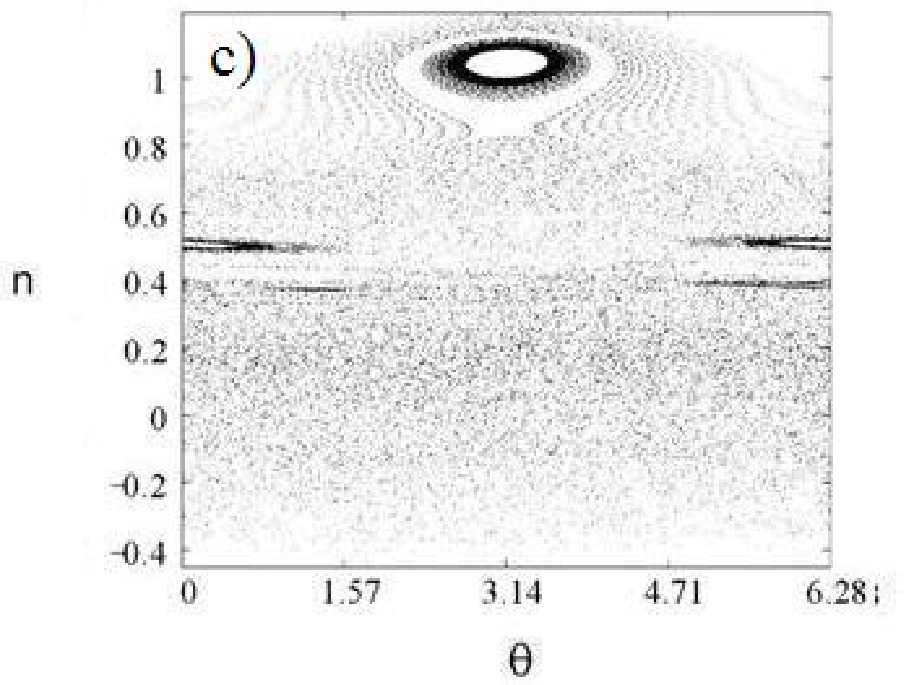}

\vspace{.5cm} \caption{  Phase space portraits of the periodically
driven: a) hydrogenlike atom ; b) quarkonium ($a \gg 1$ case); c)
quarkonium ($a \ll 1$ case), for $\epsilon_{cr}/\epsilon = 0.9$.}
\end{center}
\end{figure}

\end{document}